\documentclass[reqno]{fundam}
\usepackage{url}
\newcommand{\ctr}[1]{\begin{center} #1 \end{center}}

\renewcommand{\`}{{`}}

\newcommand{\alx}[1]{\begin{align*} #1 \end{align*}}

\newcommand{\dom}{\mbox{dom}}

\begin{document}
\title{Very simple Chaitin machines for concrete AIT}
\author{Michael Stay\\
Department of Computer Science\\
The University of Auckland\\ 
Private Bag 92019, Auckland, New Zealand\\ 
{\tt msta039@ec.auckland.ac.nz} \\ 
}

\runninghead{M. Stay}{Very simple Chaitin machines for concrete AIT}

\maketitle

\begin{abstract}
In 1975, Chaitin introduced his celebrated Omega number, the halting probability of a universal Chaitin machine, a universal Turing machine with a prefix-free domain.  The Omega number's bits are {\em algorithmically random}---there is no reason the bits should be the way they are, if we define ``reason'' to be a computable explanation smaller than the data itself.  Since that time, only {\em two} explicit universal Chaitin machines have been proposed, both by Chaitin himself.

Concrete algorithmic information theory involves the study of particular universal Turing machines, about which one can state theorems with specific numerical bounds, rather than include terms like $O(1)$.  We present several new tiny Chaitin machines (those with a prefix-free domain) suitable for the study of concrete algorithmic information theory.  One of the machines, which we call Keraia, is a binary encoding of lambda calculus based on a curried lambda operator.  Source code is included in the appendices. 

We also give an algorithm for restricting the domain of blank-endmarker machines to a prefix-free domain over an alphabet that does not include the endmarker; this allows one to take many universal Turing machines and construct universal Chaitin machines from them.  
\end{abstract}

\section{Introduction}
In 1948, Shannon published his seminal paper on {\em information theory} \cite{Shannon}.  In Shannon's model, there is a sender, a receiver, and a (probably noisy) channel, or pipe.  The sender transmits strings of bits over the channel, and the receiver records them on arrival.  To let the receiver know when to stop listening, the message must have a certain pre-arranged structure.  The message may end with a binary equivalent of ``over and out,'' or perhaps the length of the message is sent first.  What matters is that as soon as the last bit is sent, the reciever can calculate that that bit was the end of the message; such structures are known as {\em instantaneous codes}.  

Not all instantaneous codes are created equal: some can send the same information with far fewer bits than others.  The sender and receiver can agree on a {\em dictionary} that associates common strings with short encodings and  infrequent strings with longer ones.  If the message is typical, this will save considerably on the message size.  If the message is atypical, there's more information in it, more that is unexpected.  The shortest possible message size is related to the amount of information the string contains relative to the chosen dictionary, the string's {\em complexity}.

In the mid-1960's, Kolmogorov \cite{Kolmogorov}, Solomonoff \cite{Solomonoff}, and Chaitin \cite{Chaitin66} independently proposed the idea of using programs to describe the complexity of strings.  This gave birth to the field of {\em algorithmic information theory} (AIT) \cite{Chaitin77}.  The Kolmogorov complexity $K_U(s)$ of a string $s$ is the length of the shortest program in the programming language $U$ whose output is $s$.  Solomonoff proposed a weighted sum over the programs, but it was dominated by the shortest program, so his approach and Kolmogorov's are roughly equivalent.  Chaitin added the restriction that the programs themselves must be codewords in an instantaneous code, giving rise to {\em prefix-free AIT}.  In this model, complexities become true probabilities, and Shannon's information theory applies directly.

Given a prefix-free domain, there is the natural distribution $P(x) = 2^{-|x|}$, where $|x|$ is the length of $x$ in bits.  One can then ask the question, ``What is the probability, given this distribution of inputs, that a program will output {\em some} string and halt?''  Chaitin discovered that the bits of this ``halting probability'', after an initial computable prefix, are pure information \cite{Chaitin99}: the length of the shortest program that computes the first $n$ bits of the halting probability and stops is at least $n-c$ bits long.  To calculate one more bit, you have to add at least one more bit to your program; there is no description of the strings of bits shorter than the strings themselves, modulo some fixed constant.  The bits also contain all the information about whether each program will halt or not; Chaitin called this halting probability an Omega number.

With the advent of accessible computers, Chaitin proposed studying {\em concrete} AIT---theorems about specific programming languages, with positive integers as error terms rather than the phrase ``some fixed constant.''  To study concrete prefix-free AIT, Chaitin proposed two universal languages, or machines: one was a variant of LISP \cite{Chaitin99}; the other an 11-instruction ``register machine'' \cite{Chaitin87}.  We know of no other universal Chaitin machines in the literature.  LISP is a complex language, and the register machine, while small, is far from minimal.  To foster the study of concrete AIT, we propose a few new minimalist Chaitin machines.

\section{Definitions}\label{def}
Fix an alphabet $\Sigma$.  The set of all finite strings of elements of $\Sigma$ is denoted $\Sigma^*$.  A {Turing machine} $M$ is a partial recursive function $M: \Sigma^* \rightarrow \Sigma^*$, where ``partial'' means that $M$ may be undefined on some inputs; this handles the cases where the program runs forever.  

A Turing machine $M$ is called a {\em Chaitin machine} if its domain 
\[ \dom(M) \equiv \{x \in \Sigma^* : M(x) \mbox{ halts}\} \] 
is prefix-free, i.e. for all $x,y \in \dom(M)$, either $x$ is a prefix of $y$, $y$ is a prefix of $x$, or $x$ is identical to $y$.  

For any $x \in \Sigma^*$ and Chaitin machine $M$, the program-size complexity of $x$ with respect to $M$ is defined as 
\[ H_M(x) \equiv \mbox{ min}\{|w| : M(w) = x\}.\]

A Chaitin machine $U$ is called {\em universal} with respect to a set of machines $\mathbf{S}$ if $H_U(x) \leq H_M(x) + O(1)$ for any $x \in \Sigma^*$ and for any $M \in \mathbf{S}$.

It is helpful to consider a Chaitin machine in Shannon's original sender-pipe-receiver model.  Borrowing terminology from concurrent programming, the pipe is a shared resource.  The input to the machine is held by the sender, a producer.  The sender tries to put its bits into the pipe; it blocks if there are more bits to send and the pipe is full.  When there are no more bits to send, the sender halts.  The Chaitin machine is the receiver, a consumer.  From time to time it tries to get bits out of the pipe, and blocks if the pipe is empty.  The entire computation is said to halt if the sender halts, the Chaitin machine halts, and the pipe is empty.  {\em Codewords} are those inputs $x$ for which the computation halts, {\em i.e.} $x \in \dom(M)$.

This model makes it easy to see why the domain of a Chaitin machine is prefix-free: any extension of a codeword would cause the sender to block, a condition called {\em overflow}; any prefix of it would cause the Chaitin machine to block, a condition called {\em underflow}.  A universal Chaitin machine usually reads in a self-delimiting program description, the {\em prefix}, and then simulates that program acting on the remainder of the input.  If the input does not contain a proper program description, the universal machine blocks; we call this condition a {\em syntax error}.

A {\em blank-endmarker machine} (BEM) $B$ is a Chaitin machine in which one symbol of $\Sigma$---the blank endmarker, hereafter denoted $\diamond$---is reserved.  All but the last symbol of the codeword are taken from the alphabet $(\Sigma - \{\diamond\})$, and the codeword is terminated with $\diamond$.  $B$ will request input until it reads the symbol $\diamond$, after which no more input will be requested.  It is this blank endmarker that allows the codewords to be prefix-free: removing the endmarker will cause an underflow; any symbols following an endmarker will cause an overflow.  

We may construct a BEM $M'$ from an arbitrary Turing machine $M$ defined on the alphabet $\Sigma$ as follows.  First, define the alphabet $\Sigma' \equiv \Sigma \cup \{\diamond\}$ over which $M'$ operates.  Next, define $\dom(M') \equiv \{ x\diamond | x \in \dom(M) \}$, i.e. we append the symbol $\diamond$ to each string on which  $M$ halts.  No prefix or extension of a codeword is in the domain of $M'$, since every codeword in the domain has exactly one $\diamond$ as the last symbol.

A {\em universal BEM} is a BEM defined on the alphabet $\Sigma'$ that is universal with respect to all such BEMs.

\begin{theorem} A universal BEM $B_U$ exists. \end{theorem} \begin{proof} Let $0^n$ denote the concatenation of $n$ 0 symbols. Then $B_U$ can read in in the self-delimiting prefix $0^n1$ and simulate $B_n$ on the remainder of the input. 
\end{proof}

\begin{theorem} \label{BEMnotUniv}No BEM is universal with respect to all Chaitin machines defined on $\Sigma'$.\end{theorem}
\begin{proof} For a BEM to be universal over a set $S$, it must be able to represent the domain of machines in $S$ with only a constant increase in the length of the codewords.  However, this is impossible.  

Consider the following Chaitin machine $C$: inputs are concatenations of a self-delimiting program $p$ and a string $x \in \Sigma'^*$.  The program $p$, when executed, outputs the length of the string, $n=|x|$.  The machine $C$ first reads $p$, then executes it to get $n$, and then reads $x$.  Finally, $C$ outputs the string $x$.  

The number of $n$-symbol strings $|\Sigma'|^n = (|\Sigma|+1)^n$.  On the other hand, the number of $(n+c)$-symbol strings available to a BEM is only $|\Sigma|^{n+c}$, because it may only use the symbol $\diamond$ once at the end of the codeword.  Since $(|\Sigma|+1)^n$ grows faster than $|\Sigma|^{n+c}$, there is no constant $c$ such that $|\Sigma|^{n+c} \geq (|\Sigma|+1)^n$ for all $n$. \end{proof}

We define the relation ``universal-with-respect-to'' and denote it $\succeq$.  Let $\Sigma_n \equiv \{0,1,\ldots,(n-1)\}$ and $\Sigma_n' \equiv \{0,1,\ldots,(n-1),\diamond\}$.  For all $n \geq 2$ we have the following: 

\begin{theorem}Let $A_n$ be Chaitin machine that is universal with respect to all Chaitin machines defined over $\Sigma_n$, $B_n$ be a BEM that is universal with respect to all BEMs defined over $\Sigma_n'$, and $C_n$ be a Chaitin machine that is universal with respect to all Chaitin machines defined over $\Sigma_n'$.  Then  
\begin{enumerate}
\item $C_n \succeq B_n \succeq A_n$, but \label{cba} 
\item $B_n \not \succeq C_n$ \label{bc}
\item and $A_n \not \succeq B_n$. \label{ab}
\end{enumerate}
\end{theorem}
\begin{proof} The first part of (\ref{cba}) holds because BEMs are Chaitin machines and $C_n$ was chosen to be universal with respect to that set.  

The second part of (\ref{cba}) holds because $B_n$ can simulate $A_n'$, the BEM constructed from $A_n$:  it reads a self-delimiting program for $A_n$ and begins simulating it.  If the program requests an input and $B_n$ reads $\diamond$, then $B_n$ loops forever, simulating the underflow condition.  If the program halts, then $B_n$ reads one more symbol, $x$; if $x \neq \diamond$, then $B_n$ loops forever, simulating overflow.  If the program loops forever on its own, then so does $B_n$.  Thus the domain of $B_n$ is the same as that of $A_n$ modulo the final $\diamond$.

Item (\ref{bc}) is theorem \ref{BEMnotUniv}.  

Finally, (\ref{ab}) holds because $A_n$ is not allowed to use $\diamond$.  It must use a self-delimiting description of the input string, and the shortest self-delimiting version of a string $x$ grows like $|x| + log^* |x| + O(1)$ \cite{Calude}, which violates the error bound for universality. \end{proof}

\section{Some minimalist machines}

In this section, we review four languages that greatly influenced our designs, and point out why these are not universal Chaitin machines.  

\subsection{Lambda calculus}
Lambda calculus formed the basis of Church's 1936 negative answer \cite{Church1, Church2} to Hilbert's {\em Entscheidungsproblem} (decision problem): is there an algorithm for deciding whether first-order statements are universally valid?  He showed first that stating the equivalence of two lambda terms was a first-order predicate, and then that there is no recursive (or computable) function that can compute whether two terms are equivalent.  Turing independently proved the same result the same year \cite{Turing}, and when he heard of Church's result, was quickly able to show that his machines compute the same class of functions as Church's lambda terms.

Everything in lambda calculus is a function; there are no built-in types, data structures, branching instructions, or constants, and the only operation is functional composition, or {\em application}.  Functions take functions as input and return functions as output.  To denote his functions, Church used a slightly different notation than most mathematicians are used to: rather than 
\[f(x,y,z) = \quad\langle\mbox{definition of f in terms of x,y,z}\rangle,\] 
Church wrote\footnote{Strictly speaking, even the equals operator is just syntactic sugar: lambda terms are {\em anonymous} functions.}
\[f=\lambda xyz.\langle\mbox{definition of f in terms of x,y,z}\rangle.\]  
Application is denoted by concatenation or parentheses, and is left-associative: $f=\lambda xy.y(xx)y$ is the same as $f(x,y) = (y(x(x)))(y)$.  

There is a {\em universal basis} consisting of the two functions (or {\em combinators}) 
\[S=\lambda xyz.xz(yz)\quad \mbox{and} \quad K=\lambda xy.x\]
That is, the function represented by a lambda term may also be represented by a combination of these combinators.  For example, the identity function $I=SKK$:
\alx{
SKKv &= (\lambda xyz.xz(yz))KKv\\
     &= Kv(Kv)\\
     &= (\lambda xy.x)v(Kv)\\
     &= v
}
In fact, there is an algorithm called {\em lambda abstraction} that reduces any lambda term to a combination of $S$ and $K$, and $I$ combinators\footnote{The combinator $I$ is included merely for convenience; it can, of course, be replaced by $SKK$.} that eliminates the need for any variables.  To abstract away all the variables in a term, begin with the innermost variable $v$ and apply the following rules, then repeat for the remaining variables.

\begin{enumerate}
\item $\mbox{abstract}(XY) = S(\mbox{abstract}(X))(\mbox{abstract}(Y))$
\item $\mbox{abstract}(v) = I$
\item if a term $X$ does not depend on $v$, then $\mbox{abstract}(X) = KX$
\end{enumerate}

For example, the reverse-application combinator is $\lambda xy.yx$.  Abstracting $\lambda y$ yeilds $\lambda x.SI(Kx)$; abstracting $\lambda x$ yeilds $S(K(SI))(S(KK)I)$.

A term is said to be in {\em normal form} if the variable to be applied has not been bound to a value.  For example, $\lambda xy.y$ is a normal form, but $(\lambda xy.y)S$ gets reduced to the normal form $\lambda y.y$, and $(\lambda y.y)K$ gets reduced to the normal form $K$.  Reducing a term to normal form is equivalent to a Turing machine reaching a halting state.  Some lambda terms do not have normal forms; these correspond to computations that never finish.  For example, the term $(\lambda x.xx)(\lambda x.xx)$ reduces to itself; it is the lambda-calculus equivalent of an infinite loop.

The output of a lambda calculus computation is the normal form of the term, if it exists.  Since normal forms can be enumerated  $\grave{a}$ $la$ G\"{o}del, there are bijections from normal forms to natural numbers and to binary strings.  For the purposes of this paper, where we define machines to be partial recursive functions from binary strings to binary strings, it is convenient to choose the latter.

Lambda calculus' alphabet consists of parentheses, the symbols for the lambda operator and the dot, and symbols for variables.  Though one rarely needs more than twenty-six variables, the formalism allows for subscripts; therefore, digits for subscripts and an end-of-subscript marker are also included.  Let $k$ be the number of symbols in the alphabet.

Since there infinitely many Chaitin machines that halt on each string $x$, lambda calculus needs to be able to encode an arbitrary string with only a constant overhead in order to be a universal Chaitin machine.  When a bit string increases by one symbol, the number of representable strings increases by a factor of $k$.  However, because of the well-formedness requirement that parentheses balance in a lambda term, the number of codewords with one more symbol increases by a smaller factor.  Any encoding of strings necessarily suffers from a slight expansion, and so lambda calculus fails to reach the constant overhead bound.  

There are more requirements for a well-formed lambda term that we are ignoring here.  We perform an exact analysis of a less restrictive case where we have only one parenthesis symbol and one combinator in the next section; we'll see that it, too, fails to be a universal Chaitin machine.

\subsection{Iota}
Iota \cite{BI} is a minimalist language created by Chris Barker.  The universal basis $\{S, K\}$ suffices to produce every lambda term, but it is not necessary.  There are one-combinator bases, known as {\em universal combinators}.   Iota is a very simple universal combinator, $\lambda f.fSK$, denoted $0$.  To make Iota unambiguous, there is a prefix operator, $1$, for application.  Valid programs are preorder traversals of full binary trees.  In the tables that follow, brackets $[\cdot]$ denote taking the semantics of the argument.

\ctr{\begin{tabular}{l|l}
Syntax & Semantics\\
\hline\\
$F \rightarrow 1F_0F_1$ & $[F_0]([F_1])$\\
$F \rightarrow 0$   & $\lambda f.fSK$\\
\end{tabular}}

Fokker \cite{F} proposed a different universal combinator, $\lambda f.fS(\lambda xyz.x)$, which is slightly larger, but recovers $S$ and $K$ with fewer applications.

Like lambda calculus, there will be a normal form of the Iota term if the program halts.  As before, because normal forms are denumerable, we can make a bijection between binary strings and normal forms, ordering both lexically, and output the matching string.  Another alternative, advocated by Ben Rudiak-Gould \cite{Lazy}, is to restrict the output to a subset of normal forms, such as lists of booleans.  Any program whose normal form is not in this subset is defined to output the empty string, while the list of booleans is converted directly to a binary string.

We illustrate the execution of two simple Iota programs below:

\alx{
100 &= (\lambda f.fSK)(\lambda f.fSK)   \\
    &= (\lambda f.fSK)SK \\
    &= SSKK \\
    &= SK(KK) \\
    &= I \\
    \\
1010100 &= 1010I\\
        &= (\lambda f.fSK)((\lambda f.fSK)I) \\
        &= (\lambda f.fSK)(ISK) \\
        &= (\lambda f.fSK)(SK) \label{internal} \tag{1}\\
        &= SKSK \\
        &= KK(SK) \\
        &= K 
}

Notice that at step (\ref{internal}) we performed an application within an internal branch. Iota is confluent: it does not matter in which order the applications are carried out, because there are no side-effects.

Iota is not quite a universal Chaitin machine because of the requirement that codewords be preorder traversals of a full binary tree.  There are $C_n$ full binary trees with $n+1$ leaves, where $C_n$ is the $n$th Catalan number, giving a codeword of length $2n+1$; asymptotically, $C_n \sim \frac{4^n}{\sqrt{\pi n^3}}$.  Thus, if we increase the length of a codeword by two bits, the number of representable strings only increases by $2-3 \lg(\frac{n}{n-1})$ bits.  It is asymptotically close to being a universal Chaitin machine, but doesn't quite make it.  Again, any encoding of strings within Iota will necessarily suffer from a slight expansion, and will not satisfy the $O(1)$ error requirement.

\subsection{Zot}
Zot \cite{BZ} is a continuized form of Iota.  Here, $1$ is a combinator rather than an operator; in Barker's words, it is treated ``lexically'' rather than ``syncategoremically.''  The initial continuation is the trivial one, and the current continuation is applied to each combinator in turn.  This allows the program to get access to each bit of input individually.    It also makes Zot a nice G\"{o} del numbering, since every blank-terminated binary string is a valid Zot codeword and every computable function is represented.

\ctr{\begin{tabular}{l|l}
Syntax & Semantics\\
\hline\\
$F \rightarrow FB$      & $[F]([B])$\\
$F \rightarrow \diamond$& $\lambda c.cI$\\
$B \rightarrow 0$       & $\lambda c.c(\lambda f.fSK)$\\
$B \rightarrow 1$       & $\lambda cL.L(\lambda lR.R(\lambda r.c(lr)))$\\
$B \rightarrow \odot$   & $\lambda c.c(O)(P)$, where $O=\lambda abcde.KI$\\
                        & and $P$ is an output monad.\\
\end{tabular}}

Barker also includes an operator which I have denoted $\odot$, which allows the program to interact with an output monad.  It is not strictly necessary: we can ignore the $\odot$ operator and, like Iota, consider the normal form of the current continuation to be the output.

Zot is a BEM, and therefore not a universal Chaitin machine.  

\subsection{Binary lambda calculus}
Binary lambda calculus (BLC) \cite{BLC} is a language created by John Tromp in resonse to Chaitin's claim \cite{Chaitin96} that ``Lambda calculus is even simpler and more elegant than LISP, but it's unusable. Pure lambda calculus with combinators $S$ and $K$, it's beautifully elegant, but you can't really run programs that way, they're too slow.''  Tromp noted that ``There is however nothing intrinsic to $\lambda$ calculus or CL that is slow; only such choices as Church numerals for arithmetic can be said to be slow, but one is free to do arithmetic in binary rather than in unary,'' and proposed BLC specifically for studying concrete AIT.  

Rather than follow Church's original notation, Tromp used de Bruijn \cite{deBruijn} notation, which eliminates the need to use the variable name in the both lambda prefix and in the body of a term.  Instead, $n$ refers to the variable bound by the $n$th enclosing $\lambda$.

\ctr{\begin{tabular}{l|l}
Syntax & Semantics\\
\hline\\
$F \rightarrow 01F_0F_1$  & $[F_0]([F_1])$\\
$F \rightarrow 00F$       & $\lambda [F]$\\
$F \rightarrow 1^{n+1}0$  & $n$\\
\end{tabular}}

Any remaining bits are converted to a $nil$-terminated list of combinators $K$ and $KI$ to which the program is applied.  The list is constructed using the pairing cobinator $P = \lambda xyz.zxy$, which has the property that $PXYK = X$ and $PXY(KI) = Y$.  Thus $K$ and $KI$ behave like booleans with respect to $P$.

Tromp explicitly states that the normal form of the resulting BLC term is the output.

Prefix-free BLC uses a rather nonstandard approach.  Instead of defining a computer whose domain is prefix-free, Tromp redefines the way output is handled: a program is prefix free if and only if 
\[ U(p:z)=\langle x_p,z\rangle, \tag{*} \] 
where $p:z$ is a lambda term encoding a list whose first few members are the bit string $p$, followed by the tail $z$, where $z$ may be infinite.  Since $z$ is potentially infinite and is arbitrary,  $U$ cannot output $z$ without processing all of the bits of $p$ and returning $z$ as part of the output.  This guarantees that no prefix or extension of $p$ has the right form, and thus the set of such $p$ is prefix-free.

Normal BLC is a BEM, and therefore not a universal Chaitin machine; prefix-free BLC is a BEM combined with the definition (*), but does not define a Chaitin machine whose domain matches the set $\{p \;|\; (*) \mbox{ holds}\}$.

We would like to construct universal Chaitin machines from universal BEMs.  The first step is to get rid of the blank endmarker.  In the next section, we describe an algorithm to extract the (possibly empty) subset of the domain of a BEM that does not depend on reading the blank endmarker to know when to halt.

\section{Eliminating the blank endmarker}\label{bem}
Some BEMs have the property that their behavior after the final read request does not depend on the result of that request.  In a BEM $B$, the last request is always for the $\diamond$ symbol, but in these cases, the program does not need the $\diamond$ to know when to halt.  We can use this property to define a Chaitin machine whose domain is a prefix-free set of programs such that if $x \in \dom(C)$, then $x\diamond \in \dom(B)$.

We construct $C$ in the following way:  
\begin{enumerate}
\item $C$ simulates $B$ up to the point where the first read request is made; if no read requests are made, then $C$ loops forever.  After a read request, no read is actually performed; rather, 
\item \label{read} $C$ simulates the behavior of $B$ for all the possibilites for that symbol up to the point where one of the branches is about to halt or make another read request.
\item At that point, the read for the previous symbol is actually performed; if there are no more symbols, then $C$ blocks; otherwise the $C$ selects the appropriate branch and the abandons other branches' simulations.
\item $C$ continues executing that branch until it halts or makes a read request.  If the selected branch halts, then $C$ halts (although if there are symbols remaining in the pipe, the sender will block and the computation will fail to halt).  If the selected branch performs a read request, $C$ goes to step (\ref{read}).
\end{enumerate}

In this way, $C$ only needs to simulate $|\Sigma|$ concurrent branches at a time, and if $B$ does not need to read $\diamond$ to know when to halt, then that read is never actually performed by $C$.  The domain of $C$ is prefix free, since $C$ would underflow on any string more than one symbol shorter than a codeword of $B$ and overflow on any extension of a codeword of $C$.

A Chaitin machine constructed in this way is universal if it can simulate either of the universal machines that Chaitin proposed with only an additive constant increase in the size of the input.  The machine constructed in this way from BLC is universal.

\section{Church's lambda operator as a curried interpreter} \label{curry}
In this section, we take a small aside to introduce a concept used in one of the proposed Chaitin machines below, as well as to introduce a bit of new notation.

Church's lambda operator can be seen as an interpreter.  It reads three self-delimiting parameters from the input stream---$var$, $body$, and $replacement$---builds data structures representing the application of the functions and operators that those parameters describe, and performs alpha and beta reduction on the data structures, calling itself recursively.  If the process of alpha- and beta-reduction reaches a point where it cannot continue, it decodes the data structure into the normal form of the lambda term and outputs it as a string of symbols.

Since an interpreter is merely a function, we can curry it: the function $\lambda$ takes an input $var$ and returns a new function $\lambda'$; likewise, the function $\lambda'$ takes a single input $body$ and returns a function $\lambda''$; the function $\lambda''$ applied to the input $replacement$ yeilds the normal form, if one exists.

Since curried functions take exactly one input, the application operator becomes strictly binary, and the tree of applications is a full binary tree.  Programs may be written as preorder traversals of the tree to avoid parentheses.  We adopt the backtick (`) as a prefix application operator throughout the rest of the paper, which behaves identically to Iota's $1$ operator; for example, the lambda term $S=\lambda xyz.xz(yz)$ will be written \ctr{\`\`$\lambda x$ \`\`$\lambda y$ \`\`$\lambda z$ \`\`$x$ $z$ \`$y$ $z$\\}

In Appendix \ref{LC2JS}, we include the Javascript source code for a source-to-source filter from this dialect of lambda calculus to Javascript.  It supports lazy evaluation and can be trivially modified to work with any eager language with first-class functions, such as Perl.

\section{A very simple Chaitin machine}
Below, we present a Chaitin-universal combinator for use in Iota.  Input requests $R$ go via a monad that binds the requests together in lazy-evaluation order.  $R$ reads a bit and evaluates to $K$ or \`$K$ $I$ if the bit is $0$ or $1$, respectively.

The definition of the combinator is optimized, like Fokker's, for the number of applications to recover $K, S$, and $R$.  Codewords are concatenations of programs (preorder traversals of the application tree) and (possibly empty) input.

\ctr{\begin{tabular}{rl}
$A$ &= \`$K$ \`$K$ $R$\\
$B$ &= \`$K$ \`$K$ \`$K$ \`$K$ \`$K$ \`$K$ \`$K$ $K$\\
$C$ &= \`\`$\lambda$ $x$ \`$x$ $B$\\
$0$ &= \`\`$\lambda$ $x$ \`\`\`\`$x$ $C$ $A$ \`$K$ $I$ $S$\\
\\
$100$ &= $K$\\
$10100$ &= $S$\\
$1010100$ &= $R$\\
\end{tabular}}

\ctr{\begin{tabular}{|l|l|}
\hline
$1$ & first program to loop due to\\ 
    & a syntax error\\
\hline
$00$ & first to loop due to overflow\\
$= 0$ with input 0 &\\
\hline
$110101000$ & first to loop due to underflow\\
$ = $\`$R$ 0 with no input &\\
\hline
$1101010000$ & first to halt with nonempty\\
$=$ \`$R$ 0 with input 0 & input\\
$=$ \`$K$ 0 &\\
\hline       
\end{tabular}}

This language is prefix-free.  Prefices of programs are not traversals of full binary trees, so they loop due to a syntax error.  In any halting program, there will be a finite number of applications of the $R$ operator, and thus a finite number of bits appended to the end of the program.  The execution of any prefix of that codeword will block due to underflow, and any extension will block due to overflow.

It is also universal with respect to all Chaitin machines defined over $\{0, 1\}$.  Chaitin's universal machine can be simulated by this one with a program that reads in a parenthesis-balanced LISP S-expression and evaluates it: $S$ and $K$ are universal over the lambda terms, and $R$ provides the bits for Chaitin's $readBit$ function.

\section{Extending a universal combinator}
One may also construct a Chaitin-universal combinator $0$ from any universal combinator $U$ by taking
\ctr{\begin{tabular}{rl}
$0$ &= \`\`$Pair$ \`\`$\lambda x$ \`\`$\lambda y$ \`\`$\lambda z$ $U$ $R$,\\
\end{tabular}}
where 
\ctr{\begin{tabular}{rl}
$Pair$ &= \`\`$\lambda x$ \`\`$\lambda y$ \`\`$\lambda z$ \`\`$z$ $x$ $y$.\\
\end{tabular}}

Programs written for $U$ can be converted to use $0$ by replacing $U$ with $100$.  For example, taking Iota's combinator $U=$\`\`$\lambda f$ ``$f$ $S$ $K$, we have that

\ctr{\begin{tabular}{rll}
$1100100$ &= \`$U$ $U$ &= $I$\\
$11001100100$ &= \`$U$ \`$U$ $U$ &= \`$S$ $K$\\
$110011001100100$ &= \`$U$ \`$U$ \`$U$ $U$ &= $K$\\
$1100110011001100100$ &= \`$U$ \`$U$ \`$U$ \`$U$ $U$ &= $S$\\
$1011001100100$ &= \`$0$ \`$U$ \`$U$ $U$ &= $R$\\
\end{tabular}}

\section{Keraia, continuized binary lambda calculus with a 6-bit UTM}

Keraia is a BEM that uses a straightforward encoding of the curried $\lambda$ introduced in section \ref{curry}.  We begin with three examples:

{\small
\noindent
\begin{tabular}{lllll}
\\
I = & $\diamond$&110&0&0\\
    & Interpret &\`\`$\lambda$&x&x\\
\end{tabular}
\\

\noindent
\begin{tabular}{lllllll}
K = & $\diamond$&110&0&110&10100&0\\
    & Interpret &\`\`$\lambda$&x&\`\`$\lambda$&y&x\\
\end{tabular}
\\

\noindent
\begin{tabular}{llllllllllllll}
S = &$\diamond$&{$\!\!\!$}110&{$\!\!\!$}10100&{$\!\!\!$}110&{$\!\!\!$}11000&{$\!\!\!$}110&{$\!\!\!$}0&{$\!\!\!$}11&{$\!\!\!$}10100&{$\!\!\!$}0&{$\!\!\!$}1&{$\!\!\!$}11000&{$\!\!\!$}0\\
&Interpret&{$\!\!\!$}\`\`$\lambda$&{$\!\!\!$}x&{$\!\!\!$}\`\`$\lambda$&{$\!\!\!$}y&{$\!\!\!$}\`\`$\lambda$&{$\!\!\!$}z&{$\!\!\!$}\`\`&{$\!\!\!$}x&{$\!\!\!$}z&{$\!\!\!$}`&{$\!\!\!$}y&{$\!\!\!$}z\\
\\
\end{tabular}
}

The leftmost leaf represents the curried $\lambda$ function; the first right subtree $var$ (if it exists) represents a variable; the second right subtree $body$ (if it and $var$ exist) represents the applications of curried lambda and variables; the third right subtree $replacement$ (if it and the previous two exist) represents the replacement pattern.

Keraia uses a greedy algorithm while marking variables: it traverses $body$ marking occurrences of $var$, then recursively parses $body$ to mark the rest of the variables.  Next, it performs $\alpha$-reduction and $\beta$-reduction until $body$ has reached normal form.  Any remaining leaves are replaced by the $0$ combinator.  Finally, $Keraia$ performs lambda-abstraction and returns a combinator.  

\ctr{\begin{tabular}{l|l}
Syntax & Semantics\\
\hline\\
$F \rightarrow FB$      & \`$[F]$ $[B]$\\
$F \rightarrow \diamond$& $[0]$\\
$B \rightarrow 0$       & \`\`$\lambda c$ \`$c$ $Interpret$\\
$B \rightarrow 1$       & \`\`$\lambda c$ \`\`$\lambda A$ \`$A$ \`\`$\lambda a$ \`\`$\lambda B$ \`$B$ \`\`$\lambda b$ \`$c$ \`\`$Pair$ $a$ $b$\\
\end{tabular}}

While Zot's 1 combinator applies the left branch to the right, Keraia's 1 merely $Pair$s the branches.  $Interpret$ reads in the data structure created by the $Pair$ing, and then interprets it.  The behavior of the function $Interpret$ is only specified on combinators of the form constructed by applications of the combinators 0 and 1. 

Like Zot, Keraia has a very simple self-interpreting UTM, with the same meaning: 111000 is the encoding of ``apply the identity operator to what follows.''

In Appendix \ref{Keraia}, we give Javascript source code for an implementation of Keraia that bootstraps off of the curried lambda dialect from section \ref{curry}.

\section{Keraia as a universal Chaitin machine}
Rather than use a continuized set of combinators to get a universal Turing machine, we can get a universal Chaitin machine with a few small modifications.  First, we treat $1$ lexically, interpreting the first full binary tree traversal as the program description; the remaining bits are given to the sender to push through the pipe.  Also, rather than replace remaining leaves with the $0$ combinator in the last step, we replace them with $R$ operator.  As always, syntax errors (incomplete tree traversals), overflow, and underflow cause the machine to loop indefinitely.

For example, the codeword 111010010100110001 splits into a self-delimiting program (all the bits but the last) and the input bit $1$ (the last bit).  The execution proceeds as follows:
\ctr{\begin{tabular}{rll}
\\
$11101001010011000$ &= \`\`\`$\lambda x$ \`$R$ $x$ $I$\\
&= \`$R$ $I$\\
&= \`\`$K$ $I$ $I$ & ($R\rightarrow$ \`$K$ $I$ upon reading the bit 1)\\
&= $I$\\
\\
\end{tabular}}

This modification of Keraia is a universal Chaitin machine, since every Lisp S-expression has an equivalent lambda term that is directly encodable, and the $R$ operator behaves identically to Chaitin's $readBit$ operator.

\section{Conclusion}
algorithmic information theory has much to say both about physics and philosophy.  It would be nice to experiment with tiny concrete models, but until now, there were only two programming languages that were universal Chaitin machines.  We have given examples of two new universal Chaitin machines, a modification to universal combinators that allows them to be Chaitin-universal, and an algorithm for constructing a Chaitin machine from a BEM by removing the blank-endmarker and extracting the prefix-free subset of those words.

The complexity of $n$ bits of a Chaitin Omega number is $n-c$; Calude {\em et al.} \cite{Glimpse} computed the first 64 bits of an Omega number, the halting probability of Chaitin's register machine.  We know, now, that $c$ for that machine is at least 64.  The machines proposed above are ideally suited for similar computations.

Chaitin published an exponential Diophantine equation with one parameter $n$ which has infinitely many solutions if and only if the $n$th bit of $\Omega_C$ (i.e. the halting probability of a particular universal Chaitin machine $C$) is one \cite{Chaitin87}.  These machines should make it possible to producing a smaller instance.

\section{Acknowledgements}
The author would like to thank Gregory Chaitin and John Tromp for helpful discussion, and especially Cristian Calude for his many comments on an early draft of this paper.

\appendix
\section{A lambda-calculus-to-Javascript source-to-source filter} \label{LC2JS}
\small{\begin{verbatim}
// The "Be Lazy" wrapper
function _(x){return function(){return x}}

// Evaluate the parsed source
function __(x){
    return eval("(function(){var x="+parse(x)+"; return x})()")
}

// The guts
function parse(x) {
    if (x.indexOf('`')==-1) return x;
    var pos=1, count, start, c;

    // find left tree
    for (count=0, start=pos; 
         count >= 0 && pos < x.length; 
         pos++)
    { 
        c=x.charAt(pos); 
        count += (c=='`')?1:((c==' ' || c=='^')?-1:0); 
    }
    var left = x.substring(1,pos);

    // find right tree
    for (count=0, start=pos; 
         count >= 0 && pos < x.length; 
         pos++)
    { 
        c=x.charAt(pos); 
        count += (c=='`')?1:((c==' ' || c=='^')?-1:0); 
    }
    var right = x.substring(start,pos);

    if (left.substring(0,2)=='`^')
        return "function(){return function("+
                left.substring(2)+"){return "+parse(right)+"()}}";
    return "function(){return "+parse(left)+"()("+parse(right)+")}";
}

S=__("``^x ``^y ``^z ``x z `y z");
K=__("``^x ``^y x");
I=__("``^x x");
omega=__("``^x `x x");
Omega=__("`omega omega");

// Calling syntax:
// K()(I)(Omega)(S) = S()
// or
// __("```K I Omega S")() = S()
\end{verbatim}}

\section{An implementation of Keraia} \label{Keraia}
\small{\begin{verbatim}
// Keraia's alphabet
_0=__("``^c `c Keraia");

// This version of _1 uses Javascript strings 
// instead of the Pair combinator
_1=__("``^c ``^L ````Bit L ``^c `c _(0) L "+
      "``^l ``^R ````Bit R ``^c `c _(0) R ``^r `c ``Cat l r");

// Assembles the string to interpret
Cat=function(){return 
        function(x){return 
            function(y){return 
                "1"+x()+y()
    }}};

// `Bit _0 = K, `Bit _1 = `K I
Bit=__("``^b ``b `K `K K `K `K `K `K `K `K I");

Keraia=function(){
    return function(x){
        function parse(x) {
            if (x.indexOf('1')==-1) return "_"+x+" ";

            var pos=1, count, start, c;

            // find left tree
            for (count=0;count >= 0 && pos < x.length; pos++)
            { c=x.charAt(pos); count += (c=='1'||c=='3')?1:-1; }
            var left = x.substring(1,pos);

            // find right tree
            for (count=0, start=pos; 
                 count >= 0 && pos < x.length; 
                 pos++)
            { c=x.charAt(pos); count += (c=='1'||c=='3')?1:-1; }
            var right = x.substring(start,pos);

            if (left.substring(0,2)=='10')
            {
                var arg=left.substring(2).
                        replace(/0/g,'2').
                        replace(/1/g,'3');
                return "``^_"+arg+" "+
                        parse(right.replace(
                        new RegExp(left.substring(2),'g'),arg));
            }
            return "`"+parse(left)+parse(right);
        }
        return __(parse(x())
// uncomment this line for the prefix-free version                   
//          .replace(/_0/g,'R')     
        )();
    }
}
epsilon=_0;
I_k = __("`````epsilon _1 _1 _0 _0 _0");             // ``^x x

K_k = _( epsilon()(_1)(_1)(_0) (_1)(_0)(_1)(_0)(_0)  // ``^x
                  (_1)(_1)(_0) (_0)                  // ``^y
                  (_1)(_0)(_1)(_0)(_0) );            // x
// These functions are only used in the prefix-free version
InputBits='';

// Input monad
R=function(){
    return function(x){
        // loop on underflow
        if (InputBits=='') Omega();

        var c=InputBits.charAt(0); 
        InputBits=InputBits.substring(1); 
        return c=='1'?K()(I)(x):K()(x);
    }
}

pf-Keraia=function(x){
    var pos=0, count, c;

    // find a full tree
    for (count=0; count >= 0 && pos < x.length; pos++)
    { c=x.charAt(pos); count += (c=='1'||c=='3')?1:-1; }

    // loop on syntax error
    if (count>-1) Omega();

    // Take the remainder of the bits as input
    InputBits = x.substring(pos);

    var combo = _(Keraia()(_(x.substring(0,pos))));

    // loop on overflow
    if (InputBits != '') Omega();    

    return combo;
}

// pf-Keraia("1001")() = I()
\end{verbatim}}

\end{document}